\documentstyle[12pt]{article}
\catcode`\@=11 \@addtoreset{equation}{section}  
\renewcommand{\theequation}                     
         {\arabic{section}.\arabic{equation}}   
\title{Dynamical Spacetime and the Curvature of Projective State Space}
\author{P. Leifer }
\date{School of Physics and Astronomy \\ Raymond and Beverly Sackler 
Faculty of Exact Sciences \\ 
Tel-Aviv University, Tel-Aviv 69978, Israel}
\begin{document}
\maketitle
\begin{abstract}
If universal quantum interaction is really connected
with the coset structure of deformations of quantum 
states then the curvature of projective Hilbert state 
space should be observable. I discuss some approach to 
the measurement of curvature-dependent values.
\end{abstract}
\vskip .2cm
\section{Introduction}
In the beautiful popular book \cite{Feynman} Feynman
discuss reflection of the light from the glass plate.
``Phenomenologically'' it may be described as a result
of reflection from the front and from the rear surfaces
of the plate.  Infact the spacetime analysis of 
amlitudes behavior shows that one should take into 
account an emissions from all electrons of the plate
(local event in spacetime has a nonlocal reason).
That is behind very simple rule of addinion of two
amplitudes there is some geometric picture 
(arc of small amplitudes) on the complex plane $C^1$.

This example gives us some hint that on the fundamental
level evolution of quantum state in the presence of 
a nonlocal spacetime interaction in  a ``field 
cloud'' of any quantum particle may have some hidden 
geometry as well.

On the fundamental level the spacetime integration
leads to major difficulties. But this is, as a matter of 
fact, the unfit task leading to different problem--many 
body problem, as we have in the case of the glass plate. 
Therefore, I think, this difficulty is merely artefack. 
If only fundamental aspect of interaction  is 
really interesting for us then we ought
to take into account not the spacetime omnipresence 
of scatterers, but rather {\bf entanglement of  internal degrees of freedom}. That is only ``scattering 
on the elementary target--the field cloud'' with 
the spatial diameter $r \sim 10^{-13} cm $ 
and the ``defreezing'' of internal degrees of 
freedom due to this ``scattering'', one should
looking for.
 
This ``entanglement of 
internal degrees of freedom'' infact has a geometric 
character but spacetime analysis is so restrictive 
that can not include specific rules for different 
kinds of fundamental interaction of elementary 
particles. This geometry is unacceptable 
for unification of quantum interaction and, 
therefore, for the consistent
foundation of quantum mechanics as well. However the 
unitary geometry of the projective Hilbert state space 
paves the way to some general approach to the dynamics 
of quantum states \cite{Le1,Le2,Le3,Le4,Le5,Le6}.

I argue that geometry of the projective Hilbert space
or,--maybe better--some special case of the K\"ahlerian
geometry, has from the physical point of view a 
dynamical meaning, namely: {\bf Fubini-Study
metric induces quantum universal interaction
due to the positive holomorphic sectional curvature}.
This point of view essentially differs from
the statistical interpretation of this metric, say 
\cite{CMP,Feng,AshSch,Apati}. These differences are as follwes:

A. Quantum mechanics must be based on the natural (robust)
results of QFT and symmetries of ``elementary particles''.
It means that primordial quantum numbers (integrals of 
motion) like electric charge, spin, color, beauty, etc, 
are only ``rotated charges'' and  entanglement of their        amounts ``shapes'' states of ``elementary particles''.

B. The trial process of ``shaping'' of states of 
``elmentary particles'' should be  self-consistent
because the changing of numbers of ``rotated charges'' 
(creation and decay of integrals of motion) have 
dynamical character \cite{TFD}.  Therefore the trial 
choice of a basis in Hilbert space and the stationary 
choice of the superposition of the basis states
can not be identified  with an ``elementary particle'' themselves. 
 
C. Since a priori we know neither correct vacuum state 
nor appropriate set of an immanent dynamical variables
related to conservation and deformation of this vacuum state, one should use a {\bf local trial  variables}.
The deformations of a superposition state 
of charges have coset structure \cite{Le3,Le4,Le5,Le6}.
Therefore they may be labeled  by the points of the projective Hilbert space $CP(N)$ with Fubini-Study 
metric  which defines a fundamental interactions between charges. Local dynamical variables shape a moving
frame and some of them look like creation-annihilation 
of ``elementary particles''.

D. The holomorphic sectional curvature of $CP(N)$
is identified with the intensity of fundamental 
interaction constant (fine structure constant, for instance)
not with the inverse Planck constant ($\kappa=2/\hbar$)
(to compare with \cite{CMP}, for example).

E. In this theory there is a natural affine connection
which expresses as a function of the metric tensor
of Fubini-Study. Therefore not the state vector itself 
(in accordance with the ideology of Berry-Aharonov-
Anandan \cite{Berry,AA}) subjected to comparison by the parallel transport, but those tangent
vector fields on the state manifold
that take the place of dynamical variables.

F. Spacetime structure is a derivable entity. All
paradoxical results like ``fasten-than-light-telegraph''
\cite{Gisin} or ``Everett phone'' of \cite{Polch}
are rooded in the nonadequacy of assumptions about
relationships between nonlinear quantum dynamics itself
and their spacetime  presentation.

\subsection{Affine Connection in CP(N), Setup Agreement 
and non-Abelian Gauge Theory}
I will try show that the root of difficulties
in interpretation of both ordinary (linear) quantum mechanics and its nonlinear generalizatin 
\cite{Weinberg} is the neglect of general properties of 
the comparison  procedure of quantum dynamical variables.

The problem of the comparison of quantum states
is not trivial one. As a matter of fact this lies 
in the basis  of the measuremet problem in quantum 
mechanics and closely connected with the EPR problem \cite{Yuval}. 
Let me use some passage from the article of Gisin \cite{Gisin}.
`The experimental testing quantum mechanics against 
local hidden variables do not only violate the Bell 
inequality, but they also agree remarkably well
with quantum mechanics. This supports the clime that 
if one spin of a singlet state pair is ``found'' to be
in the up state, then the other spin is in the down 
state, for the {\it same direction}' (it is my
italization P.L.). The question is: what is 
`same direction'? This is the crucial point because
this notion should have a physical meaning 
\cite{Yuval,Le3,Le4,Le5}.
The comparison of `z-direction' at A and B is, as
a matter of fact, the comparison of directions
of physical fields. Since fields have indefinite
numbers of degrees of freedom, a ``parallel 
transport'' has to be done in the projective Hilbert 
state \cite{Le4,Le5}. That is our {\it credo} 
in some ``a priori spacetime geometry'' must  be 
subjected to verification and just result of such 
quantum measurement gives us a possibility  to judge 
whether this is the ``same direction'' or not.
Futhermore, we have not any a priori geometry of 
spacetime and should constuct it basing on quantum 
setup \cite{Ash}.

Now we will descuss the procedure of the comparison
of local (in $CP(N)$) dynamical variables.
Let us assume we have the two separeted in ordinary (spacetime) sense setups A, B (like spectrum
analizer of NMR or detectors, say, $K$-mesons). Their spacetime separation
has explicit exprssion in internal (quantum)
terms and they will be describe a little bit later.
I will describe quantum dynamics 
of ``spin'' $S=\frac{N-1}{2}$ states in terms of 
relative amplitudes $\Pi^i_A$ and $\Pi^i_B$.
In this case the $CP(N)$ projective Hilbert space
takes the place of the base manifold of the tangent
fiber bundle \cite{Le3,Le4}. If we have different
states of ``spins'' in A setup and B setup, then we
have $\Pi^i_A \neq \Pi^i_B$. But even if one has
the coinsidence of the ``spin states'' he has not
degeneration since in our scheme A and B
are not merely labels. They are sets of the 
{\it physically distinguishable parameters}
$\{A\}=\{U_A(1)\times U_A(N),SU_A(N+1)/S[U_A(1)\times U_A(N)\}$, 
$\{B\}=\{U_B(1)\times U_B(N),SU_B(N+1)/S[U_B(1)\times U_B(N)\}$
in fibers and coset
transformations in the base manifold $CP(N)$. Therefore if 
even $\Pi^i_A = \Pi^i_B$, this means that
one has {\it different polarizatios in the same fiber
over general  $\Pi^i$ because one should to compare
dynamical variables and this procedure is possible
only after parallel transport these dynamical
variables in, say, A setup}. Of course, {\bf a priori}
there is no any physical connection between
relative amplitudes $\Pi^i_A$ and $\Pi^i_B$.
But a physical experience says us that a quantum 
transition in  the setup A {\bf may} induce a quantum 
transition in the setup B by some physical gauge field
transfering  an interaction. {\it In our case this 
interaction related to deformation of quantum state}
\cite{Le4,Le5,Le6}. This is the problem of ``internal 
quantum dynamics'' and it should be solved now in 
the internal sense of ``$\{A\}-\{B\}$ spacetime 
separation''. 

In both special and general relativity the clock synchronization is an important procedure.
In our case we should {\it agree of setups}
$\{A\}$ and $\{B\}$. This process includes the choice
of ``vacuum state'' $|\Psi^a>$, for example, and 
the choice of the ``axis of quantization''--
direction of the field for {\it the creation
of inversion} (field along the $Z_A$, for example).
Of course, in the $\{B\}$ setup one can choose 
different ``vacuum state'' $|\Psi^b>$ and direction
of the field along, say, $X_B$. Then the relative 
amplitudes should be calculated in some single chart, 
say, in the chart $U_a: \Psi^a \neq 0 $, where one has
\begin{equation}
\Pi^i_{(a)}=W^b_a \Pi^i_{(b)},
\end{equation}
where  $W^b_a=\Pi^b_{(a)}$,  and difference in 
the field directions should be taken into account 
under the comparison of the tangent vector fields 
over $CP(N)$.

This means that in the framework of my model I 
intend to use the comparison of not quantum states 
(rays) themselves \cite{Berry,AA} but dynamical 
variables which 
correspond their deformations \cite{Le4,Le5,Le6}.
Therefore  the natural connection in $CP(N)$ 
\begin{equation}
\Gamma^i_{kl} = -2 (\delta^i_k \Pi^{l*} + \delta^i_l \Pi^{k*})(R^2 + \sum_s^{N} |\Pi^s|^2)^{-1}
\label{connection} 
\end{equation}
corresponding to the Fubini-Study metric (\ref{metric}) plays an important role in the process 
of the comparison of these dynamical variables.
How the gauge field in ``reference Minkowski spacetime''
may arise under the local ``gauge transformations'' of
the functional frame has been shown in \cite{Le4}.
It is akin the non-Abelian gauge potential of
Wilczek-Zee \cite{Wilczek}.
Namely, we have shown that the connection (\ref{connection}) determines 
the natural intrinsic gauge potential of a local frame rotation in a tangent space of $CP(N)$ and, therefore, modification of field dynamical variables. Relationships between the Goldsone and Higgs modes arise in an 
absolutely natural way.

Now I will build the tangent fiber bundle over $CP(N)$ 
related to the process of the comparison of dynamical
variables arising at two quantum transitions (events).
For a simplicity we will compare dynamical variables 
associated with the transition in the ``vacuum'' state 
\begin{equation}
|\Psi_0>= \left(
\matrix{
e^{i\omega(\Psi)} \sqrt{\sum_{a=0}^N |\Psi^a|^2} \cr
0 \cr
. \cr
. \cr
. \cr
0 
}
\right )= R e^{i\omega(\Psi)}\left(
\matrix{
1 \cr
0 \cr
. \cr
. \cr
. \cr
0 
}
\right ),
\label{vac} 
\end{equation}
and  dynamical variables associated with the transition 
in the state 
\begin{equation}
|\Psi(f^1,...,f^N;\tau)>=R e^{i\omega(\Psi)}\left(
\matrix{
\cos\Theta \cr
\frac{f^1}{g}\sin\Theta \cr
. \cr
. \cr
. \cr
\frac{f^{N}}{g}\sin\Theta
}
\right)
\label{state}
\end{equation}
which belongs to the 
geodesic emitted from the ``vacuum'' state.
It is know that this geodesic is generated by the 
unitary matrix
$\hat{T}(\tau,g)=\exp(i\tau\hat{B})=$
\begin{equation}
\left(
\matrix{
\cos\Theta&\frac{-f^{1*}}{g} \sin\Theta&.&.&.&\frac{-f^{N*}}{g}\sin\Theta \cr
\frac{f^1}{g}\sin\Theta&1+[\frac{|f^1|}{g}]^2 (\cos\Theta -1)&.&.&.& \frac{f^1 f^{N*}}{g^2}(\cos\Theta-1) \cr
.&.&.&.&.&.\cr
.&.&.&.&.&.\cr
.&.&.&.&.&.\cr
\frac{f^{N}}{g}\sin\Theta&\frac{f^{1*} f^{N}}{g^2}
(\cos\Theta-1)&.&.&.&1+[\frac{|f^{N}|}{g}]^2 (\cos\Theta -1)
}
\right),
\label{flow}
\end{equation}
where $g=\sqrt{\sum_{k=1}^{N}|f^k|^2},\Theta=g\tau$ 
\cite{Le4,Le5}.
It is clear that in the framework of the map 
$U_0:\Psi^0 \neq 0$ all states with the norm $R$
may be spaned
by a geodesic of $CP(N)$ emitted from $(0,...,0)$ corresponding (\ref{vac}).
Now we have to have local dynamical variables
subjected to parallel transport along this geodesic.
In the linear fundamental representation of the 
action of $SU(N+1)$ one has
\begin{equation}
|\Psi(s)>=\exp(-\frac{i}{\hbar}s \hat{P}) |\Psi>,
\end{equation}
where $\hat{P},...,\hat{Q}\in AlgSU(N+1)$ are 
``polarization operators'' 
$\hat{P}=\mu H^{\sigma} \hat{\lambda}_{\sigma} \in AlgSU(N+1)$ which depend on external ``multipole 
magnetic'' or ``gluon'' field $H^{\sigma}$,
$1 \leq \sigma \leq N^2+2N$ and does not depend 
on the state of the quantum system. Under an
appropriate choice of units, $s$ is a proper time
of the setup B.  
Since  $\hat{\lambda}_{\sigma}$ matrices are ``global'',  they give the illusion of the 
omnipresence of a ``spin''  degrees of freedom.
But in the  nonlinear representation 
(realization) of the group symmetry the infinitesimal operators of the transformations depend 
on the state  and thereby local dynamical variables
are not separable from the state. Then a real compound
system will be in a self-consistent state.
This property demolishes any ({\bf real, of course, not gedanken!}) attempts  to combine a compound system in a priori chosen  states. Ordinary
quantum ideology accepts this possibility and this
leads to EPR paradox in linear quantum mecanics
and to difficulties in its nonlinear versions.
My theory seems to be very ``rigid'' construction which
evidently contradicts our experience.
{\bf One can avoid this contradiction assuming that
we have not spacetime  background on quantum level
with a priori structure}. A new construction
of spacetime we will discuss in the paragraph 2.

Returning to the fiber bundle, we should
obtain a ``point'' of the tangent bundle corresponding
$\tau$. Here $\tau$ is the parameter of action
which takes the place of the ``universal time''
of Horwitz \cite{Horw}.  In order to do it one 
must parallel transport
a tangent space from (\ref{state}) to (\ref{vac}).
As a matter of fact we should parallel transport
of rates of a state vector  changing
\begin{equation}
|v(s)>=-(i/\hbar)\hat{P}|\Psi (s)>.
\label{rate}
\end{equation}
The ``descent'' of the vector field $|v(s)>$ onto the 
base manifold 
$CP(N)$ is a mapping by the two formulas:
$f:\cal H \rm \to CP(N)$  
\begin{eqnarray}
f:(\Psi^0,...,\Psi^i,...,\Psi^N) \to (R\frac{\Psi^1}
{\Psi^0},..., R\frac{\Psi^N}{\Psi^0},...)=
(\Pi^1,...,\Pi^N),
\label{map1}
\end{eqnarray}
and
\begin{eqnarray}
\vec \xi=f_{*(\Psi^0,..., \Psi^N)} |v(s)> 
=\frac{d}{ds}(R \frac{\Psi^1}{\Psi^0},..., R \frac{\Psi^N}{\Psi^0})\Bigl|_0 \cr  
=\frac{d}{ds}(\Pi^1,...,\Pi^N)\Bigl|_0
=-(i/\hbar)[R P^1_0-P^0_0 \Pi^1+(P^1_k-(1/R)P^0_k \Pi^1)\Pi^k,...,\cr
R P^N_0-P^0_0 \Pi^N +(P^N_k-(1/R) P^0_k \Pi^N)\Pi^k].  
\label{map2}
\end{eqnarray}
The {\it restriction} of these mappings onto the
our geodesic is interesting for us.
Now after a small shift along geodesic we should 
``lift'' the new tangent
vector $\xi^i + \Delta \xi^i$ into the original Hilbert space $\cal H \rm$,
that is, one needs to realize two inverse mappings: 
$f^{-1}:CP(N) \to \cal H \rm $ at point $\Pi^i+
\Delta \Pi^i$ by the formula 
\begin{equation}
\Psi^{'0}=\frac{R^2}{\sqrt{\sum_{s=1}^N |\Pi^s+\Delta \Pi^s|^2+R^2}},...,
\quad 
\Psi^{'i}=(\Pi^i+\Delta \Pi^i) \frac{R}{\sqrt{\sum_{s=1}^N |\Pi^s+\Delta \Pi^s|^2+R^2}}.
\label{PsiPi'},
\end{equation}
or in the first approxipation
\begin{eqnarray}
f^{-1}:(\Pi^1+\Delta \Pi^1,...,\Pi^N + \Delta \Pi^N) 
\to [\Psi^0+\frac{\partial \Psi^0}{\partial \Pi^i} 
\Delta \Pi^i ,...,
\Psi^N+\frac{\partial \Psi^N}{\partial \Pi^i}\Delta \Pi^i].
\label{map-1}
\end{eqnarray}
and then
\begin{eqnarray}
f^{-1}_{* \Pi+\delta \Pi}(\vec \xi + \Delta \vec \xi)
=[v^0+\Delta v^0,v^1+\Delta v^1,...,v^N+\Delta v^N] \cr
=[\frac{\partial \Psi^0}{\partial \Pi^i}(\xi^i+
\Delta \xi^i),
 \frac{\partial \Psi^1}{\partial \Pi^i}(\xi^i+\Delta \xi^i) ,...,
\frac{\partial \Psi^N}{\partial \Pi^i}(\xi^i+\Delta \xi^i)].
\label{map-2}
\end{eqnarray}
It is may be shown that under the parallel transport
of the  $\vec \xi$ along a smooth curve, one has
\begin{equation} 
\Delta \xi^i=\xi^i(\tau)-\xi^i(0)=-\int_0^\tau \Gamma^i_{kl}
\xi^l \frac{d\Pi^k}{dl}dl,
\end{equation}
and, therefore, in the first approximation
\begin{equation} 
|\delta v(\tau)>=-\Gamma^i_{kl}(\tau)\xi^l (\tau)\delta \Pi^k \frac{\partial \Psi^a(\tau)}{\partial \Pi^i}|a>.
\end{equation}
This evolution {\it effectivly defines the map of
the local vector field of dynamical variables} $\xi^i$ in 
CP(N)  to the dynamically shifted states 
$|\Psi +\Delta \Psi>$ in 
original Hilbert space just along a geodesic
(section of bundle). 

Let me now to compare two dynamical variables
\begin{equation}
D_\sigma(\hat{P})=\Phi^i_\sigma (\Pi,P)\frac{\delta}{\delta \Pi^i}
+\Phi^{i*}_\sigma (\Pi,P)\frac{\delta}{\delta \Pi^{i*}},
\end{equation}
and
\begin{equation}
D_\sigma(\hat{Q})=\Phi^i_\sigma (\Pi,Q)\frac{\delta}{\delta \Pi^i}
+\Phi^{i*}_\sigma (\Pi,Q)\frac{\delta}{\delta \Pi^{i*}},
\end{equation}
(tangent vector fields) corresponding two quantum transitions in different
quantum states by the parallel transport in the ``vacuum'' state. Here
\begin{equation}
\Phi_{\sigma}^i(\Pi;P) = R\lim_{\epsilon \to 0} \epsilon^{-1}
\biggl\{\frac{[\exp(i\epsilon P_{\sigma})]_m^i \Psi^m}{[\exp(i \epsilon P_{\sigma})]_m^k
\Psi^m }-\frac{\Psi^i}{\Psi^k} \biggr\}=
\lim_{\epsilon \to 0} \epsilon^{-1} \{ \Pi^i(\epsilon P_{\sigma}) -\Pi^i \}
\end{equation}
are the local (in $CP(N)$) state-dependent components  of the $SU(N+1)$ group generators, which are studied in \cite{Le1,Le2,Le3}.
The connection between $\Phi^i_\sigma (\Pi,P)$ and $\xi^i$
is simply $\frac{d\Pi^i}{ds}=\xi^i=
\Phi^i_\sigma (\Pi,P)\omega^\sigma=
\mu \Phi^i_\sigma (\Pi,P) H^\sigma $.

It is very important that there are transformations
from the isotropy group of the ``vacuum'' state
that leave (\ref{vac}) intact but rotate
geodesic spanning (\ref{vac}) and (\ref{state}).
{\bf This fact seems to be paves the  way to the introduction of dynamical spacetime of the ordinary
dimension--4, since gives a possibility to truncate
the multilevel amplitudes of traversal of the geodesic
up to two-level}. In the general case geodesic obeys 
equations 
\begin{equation} 
\frac{d^2 \Pi^i}{dl^2}+\Gamma^i_{km}\frac{d\Pi^k}{dl}
\frac{d\Pi^m}{dl}=0, c.c.,
\label{gN}
\end{equation} 
which in particular case CP(1) is simply
\begin{equation} 
\frac{d^2 \Pi}{dl^2}-\frac{2\Pi^*}{R^2+|\Pi|^2}(\frac{d\Pi}{dl})^2, c.c.
\label{g1}
\end{equation} 
with the solution
\begin{equation} 
\Pi(l)=R e^{i\alpha} \tan (l).
\end{equation} 
One can render the solution of general equation
(\ref{gN}) into solution of (\ref{g1}) by ansatz
of the ``squeezing'' of full state vector (\ref{state})
to the ``two-level state'' as followes.
The first ``squeezing'' unitary matrix is
\begin{equation}
\hat{G}_1^+= \left(
\matrix{
1&0&0&.&.&.&0 \cr
0&1&0&.&.&.&0 \cr
.&.&.&.&.&.&. \cr
.&.&.&.&.&.&. \cr
0&.&.&.&1&0&0 \cr
.&.&.&.&0&cos \phi_1&e^{i\alpha_1} sin \phi_1 \cr
0&0&.&.&0&-e^{-i\alpha_1}sin \phi_1&cos \phi_1 
}
\right ). 
\end{equation}
This matrix acts on the state vector (\ref{state})
with the result
\begin{equation}
\hat{G}_1^+|\Psi>= \left(
\matrix{
\cos \Theta \cr
\frac{f^1}{g} \sin \Theta \cr
. \cr
. \cr
. \cr
\frac{f^{N-1}}{g} \sin \Theta \cos \phi_1
+ \frac{f^{N}}{g} \sin \Theta e^{i\alpha_1} \sin \phi_1 \cr
-\frac{f^{N-1}}{g} \sin \Theta e^{-i\alpha_1}\sin \phi_1
+ \frac{f^{N}}{g} \sin \Theta \cos \phi_1 
}
\right ).
\label{G1} 
\end{equation}
Now one has solve two ``equations of annihilation'' 
\cite{Le1}
$\Re (-\frac{f^{N-1}}{g} \sin \Theta e^{-i\alpha_1}\sin \phi_1
+ \frac{f^{N}}{g} \sin \Theta \cos \phi_1 )=0$
and 
$\Im ( -\frac{f^{N-1}}{g} \sin \Theta e^{-i\alpha_1}\sin \phi_1
+ \frac{f^{N}}{g} \sin \Theta \cos \phi_1)=0$ in order to
eliminate the last string and to find $\alpha'_1$ and $\phi'_1$.
That is one will have a squeezed state vector
\begin{equation}
\hat{G}_1^+|\Psi>= \left(
\matrix{
\cos \Theta \cr
\frac{f^1}{g} \sin \Theta \cr
. \cr
. \cr
. \cr
\frac{f^{N-1}}{g} \sin \Theta \cos \phi'_1
+ \frac{f^{N}}{g} \sin \Theta e^{i\alpha'_1}\sin \phi'_1\cr
0
}
\right ).
\label{G2} 
\end{equation}
The next step is the action of the matrix with the shifted
transformation block
\begin{equation}
\hat{G}_2^+= \left(
\matrix{
1&0&0&.&.&.&0 \cr
0&1&0&.&.&.&0 \cr
.&.&.&.&.&.&. \cr
0&.&.&.&1&0&0 \cr
.&.&.&.&0&cos \phi_2&e^{i\alpha_2} sin \phi_2 \cr
0&0&.&.&0&-e^{-i\alpha_2}sin \phi_2&cos \phi_2 \cr
0&.&.&.&0&0&1 
}
\right )
\label{G3} 
\end{equation}
on the vector (\ref{G2}) and the evaluation of
$\alpha'_2$ and $\phi'_2$ and so on till the initial 
vector (\ref{state}) will be reduced to the following 
form 
\begin{equation}
|F(f^1,...,f^N;\tau)>=R e^{i\{\omega(\Psi)+ \epsilon(f^1,...,f^N)\}}                 
\left(
\matrix{
\cos\Theta \cr
\sin\Theta \cr
0 \cr
. \cr
. \cr
. \cr
0
}
\right)
\label{geod}
\end{equation}
That is $|F(f^1,...,f^N;\tau)>=\hat{G}^{-1}|\Psi>$, where 
$\hat{G}= \hat{G}_1 \hat{G}_2...\hat{G}_N$.
It is easy to see that the functions 
$(\Pi(\Theta),0,...,0)$ being substituted into (\ref{gN})
where
\begin{equation} 
\Pi(\Theta)= R e^{i\{\omega(\Psi)+ \epsilon(f^1,...,f^N)\}}
\tan (\Theta)
\end{equation} 
is solution of the equations (\ref{gN}). This set is 
reduced now to the single equation 
\begin{equation} 
\frac{d^2 \Pi(\Theta)}{d\Theta^2}-\frac{2\Pi^*(\Theta)}{R^2+|\Pi(\Theta)|^2}
(\frac{d\Pi(\Theta)}{d\Theta})^2, c.c.
\label{g}
\end{equation}
This reduction to the single complex local coordinate
$\Pi(\Theta)$ which accumulated a full information
about initial and finite multilevel states is
basis for {\bf integration over dynamical spacetime 
in order to take into account entanglement of
internal degrees of freedom which have been mentioned
above}.
The ``direct'' comparison of the rates of quantum transitions is possible only in the 
original Hilbert space by {\bf the compensation of the 
geodesic shift with the help of rotations of the 
funtional frame} $\{|a>\}$. This is equivalent
to the variation of the ``multipole magnetic'' or 
``gluon'' field 
$H^\sigma \to (H+\delta H)^\sigma$ in order to
reduce to zero the difference between parallel 
transported  from (\ref{state}) to (\ref{vac}) dynamical
variables $\xi$ and dynamical variables at the ``vacuum''
state (zero method of measurement which gives explicit
answer ``yes'' or ``no''). 
Here one finds that this variation is
\begin{eqnarray}
\delta H= \frac{1}{\mu}\delta U=\frac{1}{\mu} A_m \delta \Pi^m 
=\frac{1}{\mu}\frac{\delta U}{\delta \Pi^m}\delta \Pi^m
=-\frac{\hbar}{\mu} \Gamma^i_{km}\xi^k \frac{\partial \Psi^a}{\partial \Pi^i}\delta \Pi^m |a>
\label{Am} 
\end{eqnarray}
may be  connected with the  ``instantaneous'' 
self-interacting potential associated with 
the infinitesimal gauge transformation of the local 
frame  with the coefficients (\ref{connection})
\cite{Le6}. Dynamical description of this gauge
field requires the ``internal'' intoduction of 
spacetime coordinates in pure quantum manner.

\section{Introduction of Dynamical Spacetime}
In my previous works the ``reference Minkowski 
spacetime'' has been introduced in order
to connect the ``internal dynamics'' of the relative
Fourier components of relativistic scalar field
and their spacetime propagation \cite{Le4,Le5,Le6}.
This approach is not, of course, logically consistent
because my claim is ``to forget about spacetime priority''.
It means that quantum state contains in some sense a dynamical spacetime position of transition (event).  
Therefore dynamical spacetime coordinates ought to be
the functions of relative amplitudes $\Pi^i$.
Note that a  necessity to build physics in the 
absence of a background spacetime geometry has already 
been discussed (see \cite{Ash} and bibliography 
therein).

Now I intend to introduce {\bf dynamical spacetime} which
is based on the method of the ``logical spin 1/2''.
Breafly this method was mentioned at the end of the 
Section 6 of my article \cite{Le4}. 

In order to unify quantum theory and relativity
we have to have physical elements appropriate in both
these theories. {\bf Event is undefinable primordial
element in both special (SR) and general relativity (GR) 
\cite{Einstein} but in quantum 
case an event is a quantum transition}. In SR and GR event
is point of spacetime. In quantum theory a transition
is not already pointwise element but it may be
represented as the infinitesimal deformation of quantum
generalized coherent states in projective Hilbert space 
by coset generators \cite{Le1,Le3,Le5}.  
In SR and GR one can not
say about absolute time interval or spatial length
between events but they are fundamental notions
in the framework of the theory. In qunatum theory we
have quite different situation: some pure quantum
variables are fundamental notion and spacetime 
interval is a derivable entity. I think such the most
appropriate variable is (observable) the dipole 
moment of transition, related to spatial length 
by a simplest but not unique
way $\overrightarrow{d}=e \overrightarrow{x}$.
It is well known that in atomic physics this
dipole moment may be expressed, say,  in terms of
Einstein coefficients as followes
\begin{equation}
B_{if}=\frac{2\pi}{3 \hbar^2}|e x_{if}|^2.
\end{equation}
The dipole moment of transition may be expressed
in terms of pure quantum relative amlitudes as well.
However the dipole moment is only part of the matrix
element of the Hamiltonian of interaction a quantum 
system with the quant of a gauge field, photon,
for example, 
$e <f|\overrightarrow{u}_k \overrightarrow{x}(\Pi)|i>$
where the matrix elements of radius-vector
\begin{equation}
<f|x+iy|i>=\int_0^\infty dr r^3 \phi_i\phi_f
\oint Y^*_{l',m'} Y_{l,m} \sin \theta e^{i\phi} 
d\Omega,
\label{sphxy}
\end{equation}
\begin{equation}
<f|z|i>=\int_0^\infty dr r^3 \phi_i\phi_f
\oint Y^*_{l',m'} Y_{l,m} \cos \theta d\Omega,
\label{sphz}
\end{equation}
are expressed in the terms of spherical functions
and are used for the selection rules.
This is the consequence of 
ordinary assumptions:  the classical electrodynamical 
form of interaction energy and pseudo-Euclidean scalar product in four-dimensional spacetime. 
In such way we can obtain only the photon-like dispersion law and ordinary Lorentz group which 
conserves the light cone. {\bf The question, however, is: are these assumptions 
really correct at an arbitrary short distances i.e. 
under a deep inelastic interaction?
Metric of the dynamical spacetime certainly
depends on physical conditions of setup 
at the quantun level, i.e. the correspondence between 
these forms in spacetime and vectors of energy-momentum 
(spacetime metric) \cite{Gravitation} and it presumably paves the way to the consistent theory of quantum  
gravity}.

One should take into account that spherical functions
depend only on two angles $\theta$ and $\phi$
corresponding a representation of spatial rotations
in complex Hilbert space. {\it But our approach is
opposite: we try to represent unitary group $SU(N+1)$
by generalized Lorentz transformations with appropriate rates. A priory all parameters of ``internal''
unitary group $SU(N+1)$ have not spacetime
sense like angles $\theta$ and $\phi$, etc., and only
after establish of the generalized Lorentz transformation
which lead to excplicit ``answer'', one can
restore spacetime picture of a pre-history
of the event-quantum transition}.

In ordinary quantum mechanics
one has the wave function $\Psi (x)=<x|\Psi>$
describing spacetime (or space)
distribution of quantum system. But what is $|x>$?
Physically it means that there is some more or less
localizable (in macroscopic spacetime scale at 
distance ``x'' from ``0'') quantum system --``detector'', which after interaction with our system may change 
its internal state. This interaction in general 
may change not only the internal state of the detector 
but the position ``x'' as well. However this is not so important for us. For us is very 
important only that {\it the changing of the quantum 
state is essential fact, rather than any spacetime
fixing of events}. In some sense we can, however,
to ascribe to a quantum transition (event) ``spacetime
coordinates'' $x,y,z,ct$. 
This procedure is based upon 
{\bf the identification of the coinsidence of ``answers'' (``yes''=$|1>$, ``no''=$|0>$) on ``quantum
question''}. 
We must take into account this fact by introduction the
space of coherent states 
$|\alpha>=\alpha^1|1>+\alpha^0|0>$ 
of the ``logical spin 1/2'' as a two-level system 
in the basis $\{|1>,|0>\}$.
Then spinor $(\alpha^0, \alpha^1)$ defines a point
$\pi=\frac{\alpha^1}{\alpha^0}$ of the space of coherent states $CP(1)$. 
Therefore, under sharp tuning of the setup one can  ascribe  
effective vector of polarization (dipole moment of transition) 
\begin{eqnarray}
P_1(\pi)=\frac{x}{ct}=\frac{\pi + \pi^*}{1+|\pi|^2} \cr
P_2(\pi)=\frac{y}{ct}=-i \frac{\pi - \pi^*}{1+|\pi|^2} \cr
P_3(\pi)=\frac{z}{ct}=\frac{1 - |\pi|^2}{1+|\pi|^2}.
\label{proj} 
\end{eqnarray}
and after that corresponding dynamical spacetime 
distance in appropriate normalization
\begin{equation}
\hat{X}= \left(
\matrix{
ct+z&x-iy \cr
x+iy&ct-z \cr
}
\right ).
\label{X} 
\end{equation}
Then the unitary transformations of this coordinate matrix
\begin{equation}
\hat{X'}=\hat{L}\hat{X}\hat{L^*}
\label{L} 
\end{equation}
we will interpret as
a two-sheeted covering of {\bf Lorentz group which conserves a light cone}
\begin{equation}
\det \hat{X'}= \det \hat{L}\hat{X}\hat{L^*}=\det \hat{X}
=c^2t^2-x^2-y^2-z^2,
\label{det} 
\end{equation}
and which say how one should orient oneself own setup 
and the velocity with which it must move in order to
get an explicite answer ``yes''=$|1>$ or ``no''=$|0>$.
The formulas (\ref{proj}) are an analog of the well known matrix elements of radius-vector (\ref{sphxy}) and
(\ref{sphz}) which are expressed in terms of the relative amplitudes of transition. The connection of de-Broglie 
wave phase and their surfaces (form) in spacetime for 
motion of a ``quantum particle'' and the method of 
spacetime introduction infact has already been 
described \cite{Gravitation}. 
But we should remember that only a two-level 
approximation has been 
used and there are different degrees of freedom that
are now outside of our coherent state 
space $CP(1)$ of the ``logical spin 1/2''
which is a ``support'' of the quantum dynamical 
spacetime. That is under the ``defreezing'' of a multipole
interaction (it is possible only by taking into 
account higher three- etc.-level approximations
\cite{Ostrovskii}), description in dynamical spacetime 
is very pale.
{\bf Therefore behind Lorentz group there is more wide 
group structure. The Lorentz group is only
an ``inverse representation'' of this structure 
in four-dimensional spacetime: coset transformations in projective Hilbert space $CP(N)$ and  isotropy group transformations in fiber bundle over $CP(N)$ should 
be represented in $CP(1)$ and only after that in the
momentum or coordinate spaces. There is a very 
interesting consequence of this structure:
the magnitude of a distance in the dynamical spacetime
(defined by the dipole moment of transition) depends 
not only on relative motions of setups but on the 
dynamics of multipole moments--it may be subjected
to the analog of the Fitzgerald-Lorentz contraction 
with the increasing of quadrupole, octupole etc. 
multipole components}. This is 
some justification of title ``Superrelativity'' but 
I admit that the prefix ``super'' is misleading.

\section{Testing of the Metric Nonlinearity}
Our construction based on the assumption of
physically important role of the nonlinearity
of the curved K\"ahler state space (projective
Hilbert space $CP(N)$). 
There was attempts to find some evidence of the
nonlinearity of Weinberg's form \cite{Weinberg}.
Such kind of deviations from linearity have not 
been found \cite{Bollinger}. I propose to check
different kind of nonlinearity (metric) which
connected with the curvature of the K\"ahler
state space.  

One can express infinitesimal invariant interval
in the original Hilbert space (chord) as followes
\begin{equation}
\delta L^2= \delta_{ab}\delta \Psi^a \delta \Psi^{*b}= G_{ik*}\delta \Pi^i \delta \Pi^{*k}=\sum_a \frac{\partial \Psi^a}{\partial \Pi^i}
\frac{\partial \Psi^{*a}}{\partial \Pi^{*k}} 
\delta \Pi^i \delta \Pi^{*k}
\label{interval}
\end{equation}
\cite{Le3}. That is the generalized metric tensor of the original flat
Hilbert space in the local coordinates $\Pi$ is 
\begin{equation}
G^H_{ik*}=\sum_{a=0}^N \frac{\partial \Psi^a}{\partial \Pi^i}
\frac{\partial \Psi^{*a}}{\partial \Pi^{*k}}=
R^2 \frac{(\sum_{s=1}^N |\Pi^s|^2+R^2)\delta_{ik}-
\frac{3}{4}\Pi^{*i}\Pi^k}{(\sum_{s=1}^N |\Pi^s|^2+R^2)^2}.
\label{metric}
\end{equation}
I propose to compare in a physical experiment
the full invariant interval under
deformations $\Pi^i$ of the
initial state $|\Psi_0>$ 
$\delta L^2$ in original Hilbert space and the interval
\begin{equation}
dl^2=R^2\frac{(\sum_{s=1}^N |\Pi^s|^2+R^2)\delta_{ik}-
\Pi^{*i}\Pi^k}{(\sum_{s=1}^N |\Pi^s|^2+R^2)^2}
\delta \Pi^i \delta \Pi^{*k}
=G^P_{ik^*}\delta \Pi^i \delta \Pi^{*k}
\label{distance}
\end{equation}
in the projective Hilbert space $CP(N)$ (arc).

In accordance with our approach one should to compare
local dynamical variables i.e. tangent fields associated
with the deformations of quantum state $|\Psi>$.  
We should to compare these fields in respect with
the affine connection (\ref{connection}). 
The scalar products of two rates has the sense of
a frequensy ``correlation''  
\begin{equation}
f_P^2=\frac{1}{4\pi^2}G^P_{ik^*}\xi^i \eta^{k*}.
\label{freq1}
\end{equation}
The maximum of the difference between (\ref{freq1})
and 
\begin{equation}
f_H^2=\frac{1}{4\pi^2}G^H_{ik^*}\xi^i \eta^{k*}.
\label{freq2}
\end{equation}
lies in $|\Pi|=R $.
That is under traversing of the relarive amplitudes
one can see this difference if, of course, there is
a possibility to realize physically a comparable
dynamics in ordinary (flat) Hilbert space and
in projective Hilbert space.

\section{Discussion}
It is very interesting to proof
that 2-level restriction of whole N+1-level state
in general case leads not only to dynamical
spacetime but to the ``probability'' as well.
This maybe because there are a lot of degrees of 
freedom for the arbitrary ``orientation'' of the 
quantum setup relative to, say, vector energy-momentum 
and relative to surfaces of form in the local 
dynamical spacetime and, therefore, the process of the restoration of the pre-history of quantum event--
transition is not unique.

In order to clarify my approach to this problem I 
should make a short explanation \cite{Le1,Le2,Le5}.

This concerns a simple fact that quantum 
state of any 2-level system may be presented by points
of the $CP(1)$ or its realization as Poincar\'e sphere 
$S^2$ \cite{Berry}. Physically coordinates of each point 
of $S^2$ determine the shape of the ellipse of 
polarization and its orientation.  Any ``evolution'' of quantum state (including the changing polarization character) may be labeled by points of $CP(1)$. There 
are only two elementary kinds of the state ``evolution'': 

1. Motion of the ellipse of polarization along one 
of the parallel of latitude without deformation 
(only rotation with the shape conservation);

2. Motion of the ellipse of polarization along one of the
meridian with arbitrary strong deformations of the 
shape--from the right circuit polarization through right
elliptic, linear, left elliptic, to the left circuit
polarization.

These very well known facts closely connected with the 
invariant properties of $Z_2$-graduated algebra 
$Alg SU(2)$ and geometry of the projective Hilbert space.
{\bf I intend to generalize this picture in the case 
of $SU(N+1)$ \cite{Le1,Le2,Le5} because this generalizaton presumably paves the way to the consistent quantum 
formalism and to the ``internal'' manner of arising
spacetime from the pure internal degrees of freedom}.

That is I do not intend here to solve the Pauli 
problem \cite{Weigert}. In opposite,--my goal is 
to formulate and to solve {\it inverse ``Pauli 
problem''}, namely:

{\bf In the case of N+1-level quantum system
from the minimal set of immanent local dynamical 
variables related to the dynamical group $SU(N+1)$ 
and the coset structure $SU(N+1)/S[U(1)\times U(N)]$
of state deformation to find ``spacetime orientation''
of this system in absence of a background spacetime 
structure}.
This topic will be discuss elsewhere.
\vskip 1cm
ACKNOWLEDGEMENTS
\vskip .2cm
I sincerely thank Yuval Ne'eman and Larry Horwitz for 
useful discussions and critical notes.

\end{document}